\documentclass[prl,pra,10pt]{revtex4}%

\usepackage{graphicx}
\usepackage{dcolumn}
\usepackage{bm}

\begin{document}

\title{\textbf{Decoherence of tripartite states - a trapped ion coupled to an
optical cavity}}

\author{S. Shelly Sharma}
\email[shelly@uel.br]{}
\affiliation{Depto. de F\'{i}sica, Universidade Estadual de Londrina,
Londrina 86051-990, PR Brazil }

\author{N. K. Sharma}
\email[nsharma@uel.br]{}
\thanks{}
\affiliation{Depto. de Matem\'{a}tica, Universidade Estadual de Londrina,
Londrina 86051-990 PR, Brazil }

\author{E. de Almeida}
\email[eduardo@uel.br]{}
\thanks{}
\affiliation{Depto. de F\'{i}sica, Universidade Estadual de Londrina,
Londrina 86051-990, PR Brazil }

\author{\normalsize Corresponding author: S. Shelly Sharma (shelly@uel.br)}

\begin{abstract}
We investigate the decoherence process of three qubit system obtained by
manipulating the state of a trapped two-level ion coupled to an optical
cavity. Interaction of the ion with a resonant laser and the cavity field
tuned to red sideband of ionic vibrational motion, generates tripartite
entanglement of internal state of the ion, vibrational state of ionic center
of mass motion and the cavity field state. Non-dissipative decoherence
occurs due to entanglement of the system with the environment, modeled as a
set of noninteracting harmonic oscillators. Analytic expressions for the
state operator of tripartite composite system, the probability of generating
maximally entangled GHZ state, and the population inversion have been
obtained. Coupling to environment results in exponential decay of off diagonal
matrix elements of the state operator with time as well as a phase
decoherence of the component states. 

Numerical calculations to examine the time evolution of GHZ state generation
probability and population inversion for different values of system
environment coupling strengths are performed. Using negativity as an
entanglement measure and linear entropy as a measure of mixedness, the
entanglement dynamics of the tripartite system in the presence of
decoherence sources has been analyzed. The maximum tripartite  entanglement is found
to decrease with increase in the strength of system-environment coupling.
The negativity as well as the linear entropy as entanglement
measures give qualitatively similar results, uniquely identifying maximally
entangled and separable states of the system. For large values of 
system-environment coupling strength,
the mixed states of the composite system lying at the boundary of
entangled-separable region are reached. For these 
states the negativity and linear entropy show distinctly 
different behaviour. This can be understood by noting that whereas 
the negativity measures entanglement generating quantum correlations, the linear 
entropy measures all correlations that reduce the purity of the state.

{\textbf {Keywords}}: Trapped ions, Cavity QED, Tripartite entanglement, GHZ 
state generation, Non-Dissipative Decoherence, Negativity, Linear entropy, 
Mixed states  

\end{abstract}
\maketitle

\section{Introduction}

Controlled manipulation of quantum states and implementation of quantum
logic gates are essential elements of a quantum computer. Quantum
computation relies on nonclassical properties of qubits such as
entanglement. Entanglement dynamics is of utmost importance to quantum
communication \cite{eker91}, dense coding \cite{benn92} and quantum
teleportation \cite{benn93}, as well. Cold ions in a linear trap \cite%
{wine98} offer a promising physical system to implement quantum computation,
as each ion allows two qubit state manipulation. One of the experimental
efforts to realize multipartite entanglement involves trapping two level ion
in an optical cavity. Coupling of trapped ion to quantized field inside an
optical cavity has been successfully achieved \cite{mund03}. More than one
trapped ions can be used to construct a more complex multiple function
system in which quantum states of several trapped ions entangled with cavity
photons are manipulated in a controlled manner. A string of trapped ions in
a cavity may be used for implementing quantum gates \cite{pell95}. In
quantum networks involving cavity QED setups \cite{cira97}, quantum
information is stored and processed by ions trapped in a cavity.
De-coherence of quantum systems is one of the major hurdles to implement
quantum computers. Suppression of quantum coherence occurs due to
interactions of the quantum system with environment resulting in random and
unknown perturbations of system Hamiltonian. In a previous article \cite%
{shar103}, we proposed a scheme to generate three qubit maximally entangled
GHZ state, using trapped ion interacting with a resonant external laser and
sideband tuned single mode of a cavity field. In this
article, we investigate the decoherence process of three qubit system
obtained by manipulating the state of a two level trapped ion coupled to an
optical cavity. In the context of a qubit based on a single 
Ca$^+$ ion, Schmidt-Kaler et al \cite{schm03} have found magnetic field fluctuations 
and laser frequency fluctuations to be the major decoherence sources. 
Decoherence of motional quantum states of a trapped atom coupled to engineered 
phase and amplitude reservoirs has been measured \cite{myat00,turc00}. 
The trap frequency was changed by applying potentials to trap electrodes to 
simulate a phase reservoir. In general the environment or heat bath 
is modeled as a system of noninteraction boson modes \cite{cald83}. It 
is known that for harmonic systems a heat bath is effectively equivalent to an 
external uncolrrelated random force acting on the quantum system. As such 
the decoherence effects due to magnetic field fluctuations, laser frequency 
fluctuations, and potentials applied to elctrodes can be 
modeled as a system of noninteracting boson modes. Other important decoherence 
sources are cavity losses and spontaneous decay. The case where spontaneous decay 
effects and cavity losses become important has been examined 
by Fidio et al. \cite{fidi02}. In the present work, these two 
effects are assumed to be negligible.

As the laser-ion interaction times involved in experiments with ions trapped
in a cavity are in the micro seconds range, Markovian approximation is
likely to yield undesirable results. We consider here only adiabatic
decoherence of the system with no energy exchange between the quantum system
and environment. A detailed discussion on adiabatic decoherence has been
given by Mozyrsky and Privman in ref. \cite{priv98}. Starting from an initial
state in which the system and environment are in a separable state, the state
operator of the ion-cavity system is obtained by tracing over the
environment degrees of freedom. Coherent states are used to evaluate the
trace \cite{priv98}. The decoherence effects result in a state operator with
off diagonal matrix elements decaying with time, in the eigen basis of the
interaction Hamiltonian. The state operator is used to examine the 
decoherence effects on the probability of generating maximally entangled 
tripartite GHZ state and population inversion of two level ion. 
As the coherences decrease the system tends to a purely mixed
state of energy eigenvectors. Decoherence of tripartite entanglement is
analyzed by applying the Peres-Horodecki condition for separability \cite%
{pere96} to bipartite decompositions of the composite system. A state is
separable if the partial transpose of it's density matrix with respect to a
subsystem is positive semidefinite. An entangled state violates the positive
partial transpose (PPT) separability criterion. Negativity \cite{vida02} is
an entanglement measure based on (PPT) separability criterion. We use
Negativity to characterize the entanglement and linear entropy to measure
the mixedness of three subsystems. 

\section{Unitary evolution of isolated system}

Consider a single two-level ion trapped in a high finesse optical cavity. The
frequency of  radio frequency (Paul) trap along $x$ axis is $\nu $ and the excitation
energy of the ion is $\hbar \omega _{0}$. The ion is radiated by an external
resonant laser of frequency $\omega _{L}$ and interacts with the single
mode cavity field of frequency $\omega _{c}$ tuned to red sideband of ionic vibrational
motion. Interaction with external laser field as well as the cavity field
generates entanglement of internal states of the ion, vibrational states of
ionic center of mass, and the cavity field number state. The free 
Hamiltonian of the system is
\begin{equation}
\hat{H}_{0S}=\hbar \nu \left( \hat{a}^{\dagger }\hat{a}+\frac{1}{2}\right)
+\hbar \omega _{c}\hat{b}^{\dagger }\hat{b}+\frac{\hbar \omega _{0}}{2}%
\hat{\sigma} _{z}\quad ,  \label{01}
\end{equation}%
$\hat{a}^{\dagger }(\hat{a})$ and $\hat{b}^{\dagger }(\hat{b})$ being the
creation(destruction) operators for vibrational phonon and cavity field
photon respectively. With the ion placed close to the node of cavity field
standing wave, the interaction Hamiltonian is given by%
\begin{eqnarray}
\hat{H}_{I} &=&\hbar \Omega \lbrack \hat{\sigma} _{+}\exp \left[ i\eta _{L}(\hat{a}+\hat{%
a}^{\dagger })-i\omega _{L}t\right] +\hat{\sigma} _{-}\exp \left[ i\eta _{L}(\hat{a}+%
\hat{a}^{\dagger })+i\omega _{L}t\right]\nonumber   \\
&&+\hbar g\left( \hat{\sigma}
_{+}\hat{b}+\hat{\sigma} _{-}\hat{b}^{\dagger }\right)
{\sin }\left[ \eta _{c}(\hat{a}+\hat{a}^{\dagger })\right]  ,\label{02}  
\end{eqnarray}%
where $\Omega $ , $g$, $\eta _{L}$, $\eta _{c}$ are the Rabi frequency, the
ion-cavity coupling strength, ion-laser Lamb Dicke parameter, and ion-cavity
field Lamb Dicke parameter, respectively. We work in the Lamb Dicke regime 
that is $\eta _{L}\ll 1,$ $\eta _{c}\ll 1$. In the interaction picture 
determined by the transformation $\widehat{U}=\exp (i\hat{H}_{0S}t/\hbar )$
and rotating wave approximation, for $\omega _{L}=\omega _{0}$ and 
$\omega _{c}=\omega _{0}-\nu $, the interaction Hamiltonian reduces to 
\begin{equation}
\hat{H}_{II}=\hbar \Omega \lbrack \hat{\sigma} _{+}+\hat{\sigma} _{-}]+\hbar g{\eta _{c}}%
\left[ \hat{\sigma} _{+}\hat{b}\hat{a}+\hat{\sigma} _{-}\hat{b}^{\dagger }\hat{a}%
^{\dagger }\right] . \label{03}
\end{equation}
The unitary evolution of the isolated system in interaction picture is given
by%
\[
i\hbar \frac{\partial \Psi _{I}(t)}{\partial t}=\hat{H}_{II}\Psi _{I}(t),
\]%
where $\Psi _{I}(t)=\exp (i\hat{H}_{0S}t/\hbar )\Psi _{S}(t).$

\subsection{Basis truncation}

We expand $\hat{H}_{II}$ in the space of basis vectors  $%
\left\vert i,k,l\right\rangle $ as 
\begin{equation}
\hat{H}_{II}=%
\mathop{\displaystyle\sum}%
\limits_{i,k,l,i^{\prime },k^{\prime },l^{\prime }}\left\langle %
i^{\prime },k^{\prime
},l^{\prime }\right\vert \hat{H}_{II}\left\vert i,k,l\right\rangle
\left\vert i^{\prime },k^{\prime },l^{\prime }\right\rangle \;\left\langle
i,k,l\right\vert \text{ ,}  \label{05}
\end{equation}%
where $i=\left( g\text{ or }e\right) $ and $k,l(0,1,..,\infty), $ denote the
state of ionic vibrational motion and the cavity field number state,
respectively. We isolate a four dimensional vector space containing
the computational basis vectors $\left\vert g,m-1,n-1\right\rangle ,\left\vert
e,m-1,n-1\right\rangle ,\ \left\vert g,m,n\right\rangle $ and $\left\vert
e,m,n\right\rangle $ for a given choice of $m,n$ values. The matrix
representing the interaction Hamiltonian acting in four dimensional space is%
\begin{equation}
\hat{H}_{IS}\rightarrow \left[ 
\begin{array}{cccc}
0 & \hbar \Omega  & 0 & 0 \\ 
\hbar\Omega  & 0 & \hbar g{\eta _{c}}\sqrt{mn} & 0 \\ 
0 & \hbar g{\eta _{c}}\sqrt{mn} & 0 & \hbar \Omega  \\ 
0 & 0 & \hbar \Omega  & 0%
\end{array}%
\right].   \label{06}
\end{equation}%
The unitary transformation that diagonalizes $\hat{H}_{IS}$ is easily
obtained and yields the eigenvectors 
satisfying $\hat{H}_{IS}\Phi _{p}=E_{p}\Phi _{p},$
($p=1,4$). The computational basis vectors are related to the eigenvectors 
$\Phi _{p}$ through%
\begin{equation}
\left[ 
\begin{array}{c}
\left\vert g,m-1,n-1\right\rangle  \\ 
\left\vert e,m-1,n-1\right\rangle  \\ 
\left\vert g,m,n\right\rangle  \\ 
\left\vert e,m,n\right\rangle 
\end{array}%
\right] =\left[ 
\begin{array}{cccc}
\frac{A+B}{\sqrt{2}} & \frac{A-B}{\sqrt{2}} & \frac{A-B}{\sqrt{2}} & \frac{%
-A-B}{\sqrt{2}} \\ 
\frac{A-B}{\sqrt{2}} & \frac{-A-B}{\sqrt{2}} & \frac{A+B}{\sqrt{2}} & \frac{%
A-B}{\sqrt{2}} \\ 
\frac{B-A}{\sqrt{2}} & \frac{A+B}{\sqrt{2}} & \frac{A+B}{\sqrt{2}} & \frac{%
A-B}{\sqrt{2}} \\ 
\frac{-A-B}{\sqrt{2}} & \frac{B-A}{\sqrt{2}} & \frac{A-B}{\sqrt{2}} & \frac{%
-A-B}{\sqrt{2}}%
\end{array}%
\right] \left[ 
\begin{array}{c}
\Phi _{1} \\ 
\Phi _{2} \\ 
\Phi _{3} \\ 
\Phi _{4}%
\end{array}%
\right] , \label{07}
\end{equation}%
where $a_{mn}=\frac{1}{2}g\eta _{c}\sqrt{mn}$, $\mu _{mn}=\sqrt{%
a_{mn}^{2}+\Omega ^{2}}$, $A^{2}=(\mu _{mn}+\Omega )/4\mu _{mn}$, and $%
B^{2}=(\mu _{mn}-\Omega )/4\mu _{mn}$. The corresponding eigenvalues are $%
E_{1}=\hbar(\mu _{mn}-a_{mn})$, $E_{2}=-\hbar\left( \mu _{mn}+a_{mn}\right)$, 
$E_{3}=\hbar(\mu_{mn}+a_{mn})$, and $E_{4}=\hbar(a_{mn}-\mu _{mn})$. 

\subsection{Unitary evolution in truncated basis space}

Unitary time evolution of the system due to interaction operator $ \hat {H}%
_{IS}$ had been obtained, analytically, in
ref. \cite{shar103} for initial states $\left\vert g,m-1,n-1\right\rangle ,$ and 
$\left\vert e,m-1,n-1\right\rangle $. For interaction time $t_{p}$ such that $\mu
_{mn}t_{p}=p\pi ,\ p=1,2...,$ the initial state $\left\vert
g,m-1,n-1\right\rangle $ is found to evolve into 
\begin{equation}
\Psi_I (t_{p})=(-1)^{p}\left[ \cos (a_{mn}t_{p})\left\vert
g,m-1,n-1\right\rangle -i\sin (a_{mn}t_{p})\left\vert
e,m,n\right\rangle \right] ,  \label{eq231}
\end{equation}%
Now consider a special initial state with the ion in its ground state 
occupying the lowest energy trap state,
while the cavity is prepared in vacuum state ($m=n=1)$. Without taking 
into consideration the decoherence effects, the
ratio $\alpha =\left( \mu _{11}/a_{11}\right) $ determines the interaction
time $t_{_{p}}$ needed to generate maximally entangled tripartite GHZ state.
Leaving out an overall phase factor, for an interaction time $t_{p}$ such
that $a_{11}t_{p}=\frac{\pi }{4},\frac{5\pi }{4},...,$ ($\alpha =4)$, the
system is found to be in the state 
\begin{equation}
\Psi _{GHZ,I}^{-}(t_{p})=\frac{1}{\sqrt{2}}\left( \left\vert
g,0,0\right\rangle -i\left\vert e,1,1\right\rangle \right) ,
\label{eq232}
\end{equation}%
and for $a_{11}t_{p}=\frac{3\pi }{4},\frac{7\pi }{4},...,$ the state of the
system is 
\begin{equation}
\Psi _{GHZ,I}^{+}(t_{p})=\frac{1}{\sqrt{2}}\left( \left\vert g,0,0\right\rangle
+i\left\vert e,1,1\right\rangle \right) .  \label{eq233}
\end{equation}%
The density operator representation for these states is 
\begin{equation}
\widehat{\rho }%
_{GHZ,I}(t_{p})=\left\vert \Psi _{GHZ,I}(t_{p})\right\rangle \left\langle \Psi
_{GHZ,I}(t_{p})\right\vert . \label{eq234}
\end{equation}
Neglecting an overall phase factor, the states can be written in Schrodinger picture
as 
\begin{eqnarray*}
\Psi _{GHZ}^{-}(t_{p}) &=&\frac{1}{\sqrt{2}}\left( \left\vert g,0,0 \right\rangle
-i\exp \left(  -i 2\omega _{0}t \right)  \left\vert
e,1,1\right\rangle \right)  \label{eq235}\\
\Psi _{GHZ}^{+} (t_{p})&=&\frac{1}{\sqrt{2}}\left( \left\vert g,0,0\right\rangle
+i\exp \left(  -i 2\omega _{0}t \right) \left\vert
e,1,1\right\rangle \right). \label{eq236}
\end{eqnarray*}%

\section{Decoherence of unitary evolution in truncated space}

We notice that the interaction Hamiltonian $\hat{H}_{II}$ (Eq. (\ref{05}))
connects the model space states to states outside the model space. For the
specific case of initial state $\left\vert g,0,0\right\rangle,$ the leading
term in the probability of finding a state outside the model space varies as 
$t^{8}.$ The states outside the model space may be considered as part of the
environment states. The resulting decoherence effects can be accounted for
by coupling the model space interaction Hamiltonian to the environment
modeled as a set of oscillators. We recall that the scattering outside the
model space results in internal states linked to motional states with phonon
number $>1$, resulting in heating of the ions. In a realistic exprerimental 
set-up, the heating of ions trapped 
in a radio-frequency trap occurs due to the electric field noise from 
the trap electrodes \cite{turc200}. The two effects
can not be separated. Error in initial phonon state preparation can also
add to decoherence by additional scattering to states outside the model
space. The cavity states with number of photons $>1$, are created and
annihilated in the vector space outside the model space. In a lossy cavity
photons leaving the cavity are akin to state reduction which can contribute
to creating states in the model space. In this work we consider the
average decoherence effects arising due to the decoherence sources such as
magnetic field fluctuations, laser frequency fluctuations, potentials applied 
to elctrodes and coupling to states outside the model space. The random 
perturbations of the Hamiltonian are accounted for by coupling the model space
interaction Hamiltonian to environment modeled as a set of non-interacting 
harmonic oscillators \cite{cald83}. The phase decoherence results
in decay of entanglement needed for various computation related tasks.
In interaction picture the Hamiltonian for the system and environment, 
with $\hat{H}_{IS}$ coupled to environment is given by
\begin{equation}
\hat{H}=\hat{H}_{IS}+\sum_{k}\hbar \omega _{k}\hat{B}_{k}^{\dagger }\hat{B}%
_{k}+\hat{H}_{IS}\sum_{k}\left( g_{k}^{\ast }\hat{B}_{k}+g_{k}\hat{B}%
_{k}^{\dagger }\right) ,  \label{eq2}
\end{equation}%
where $\hat{B}_{k}^{\dagger }$ and $\hat{B}_{k}$ are bosonic creation and
destruction operators for the environment mode of frequency $\omega _k$. 
The coefficients $g_{k}$, $g_{k}^{\ast }$ are the system environment couplings.

The initial state of the system is assumed to be $\widehat{\rho }_{S}(0)$,
while the environment is in an uncorrelated state, $\prod_{k}\widehat{\theta 
}_{k}$, where $\widehat{\theta }_{k}=Z_{k}^{-1}e^{-\beta \omega _{k}\hat{B}%
_{k}^{\dagger }\hat{B}_{k}}$ and $Z_{k}=(1-e^{-\beta \omega _{k}})^{-1}.$
The assumption that the environment is in an uncorrelated state allows 
exact solvability, though no physical fundamental reason can be given for this.
However, as the initial state of the system is a doctored state, the assumption of 
separability of system environment state at $t=0$, holds.
The state operator for the system is obtained by solving 
\begin{equation}
\widehat{\rho }_{IS}(t)=Tr_{E}\left( e^{\frac{-{\rm i}\hat{H}t}{\hbar }}%
\widehat{\rho }_{S}(0)\prod_{k} \hat{\theta} _{k}e^{\frac{{\rm i}\hat{H}t}{\hbar }%
}\right) ,  \label{eq3}
\end{equation}%
where $Tr_{E}$ refers to the operation of tracing over bosonic degrees of
freedom used to model the environment. Working in the eigen basis of $\hat{H}%
_{IS}$, we may write 
\begin{eqnarray}
\widehat{\rho }_{IS}(t) &=&\sum_{i,j}Tr_{E}\left( e^{-{\rm i}t\sum_{k}\left(
\omega _{k}\hat{B}_{k}^{\dagger }\hat{B}_{k}+\frac{E_{i}}{\hbar }\left(
g_{k}^{\ast }\hat{B}_{k}+g_{k}\hat{B}_{k}^{\dagger }\right) \right)
}\prod_{k} \hat{\theta} _{k}e^{{\rm i}t\sum_{k}\left( \omega _{k}\hat{B}%
_{k}^{\dagger }\hat{B}_{k}+\frac{E_{j}}{\hbar }\left( g_{k}^{\ast }\hat{B}%
_{k}+g_{k}\hat{B}_{k}^{\dagger }\right) \right) }\right)  \nonumber \\
&&\exp \left[ \frac{-{\rm i}(E_{i}-E_{j})t}{\hbar }\right] \left\langle \Phi
_{i}\right\vert \widehat{\rho }_{S}(0)\left\vert \Phi _{j}\right\rangle
\left\vert \Phi _{i}\right\rangle \left\langle \Phi _{j}\right\vert .
\label{eq4}
\end{eqnarray}%
The trace over bosonic bath states is evaluated by using coherent states 
\cite{priv98} and the resulting state operator for the system is 
\begin{eqnarray}
\widehat{\rho }_{IS}(t) &=&\sum_{i,j}\left\langle \Phi _{i}\right\vert 
\widehat{\rho }_{S}(0)\left\vert \Phi _{j}\right\rangle \exp \left( \frac{%
-(E_{i}-E_{j})^{2}\Gamma (t)}{4.0\hbar ^{2}}\right)  \nonumber \\
&&\exp \left[ -{\rm i}\left( \frac{(E_{i}-E_{j})t}{\hbar }+\frac{%
(E_{i}^{2}-E_{j}^{2})C(t)}{\hbar ^{2}}\right) \right] \left\vert \Phi
_{i}\right\rangle \left\langle \Phi _{j}\right\vert ,  \label{eq5}
\end{eqnarray}%
where, using the notation of ref. \cite{priv98}, 
\begin{equation}
\Gamma (t)=8\sum_{k}\frac{\left\vert g_{k}\right\vert ^{2}}{\omega _{k}^{2}}%
\sin ^{2}\left( \frac{\omega _{k}t}{2}\right) \coth \left( \frac{\beta
\omega _{k}}{2}\right) ,  \label{eq6}
\end{equation}%
and 
\begin{equation}
C(t)=\sum_{k}\frac{\left\vert g_{k}\right\vert ^{2}}{\omega _{k}^{2}}\left(
\sin (\omega _{k}t)-\omega _{k}t\right) .  \label{eq7}
\end{equation}%
The summation over bath modes in Eqs. (\ref{eq6}) and (\ref{eq7}) can be
replaced by integration over frequency by introducing a frequency dependent
density function. For an ohmic dissipation characterized by density function 
$D(\omega )\left\vert g\right\vert ^{2}=\kappa \omega \exp (-\omega /\omega
_{c}),$ we obtain 
\begin{equation}
\Gamma (t)=8\kappa \int d\omega \frac{\exp (-\omega /\omega _{c})}{\omega }%
\coth \left( \frac{\beta \omega }{2}\right) \sin ^{2}\left( \frac{\omega t}{2%
}\right) ,  \label{eq8a}
\end{equation}%
and 
\begin{equation}
C(t)=\kappa \int d\omega \frac{\exp (-\omega /\omega _{c})}{\omega }\left(
\sin (\omega t)-\omega t\right) ,  \label{eq8b}
\end{equation}%
where $\omega _{c}$ is a cut off frequency and constant $\kappa $ a measure
of system-bath coupling strength.

\subsection{Initial State $\widehat{\protect\rho }_{S}(0)=\left|
g,m-1,n-1\right\rangle \left\langle g,m-1,n-1\right| $}

We consider the case when the system is prepared, at $t=0,$ in the state $%
\widehat{\rho }_{S}(0)=\left\vert g,m-1,n-1\right\rangle \left\langle
g,m-1,n-1\right\vert $. The state operator $\widehat{\rho }_{IS}(t)$
obtained by putting in the energy spectrum of $\hat{H}_{IS}$ in Eq. (\ref%
{eq5}) is 
\begin{eqnarray}
\widehat{\rho }_{IS}(t) &=&\frac{(A+B)^{2}}{2}\left[ \left\vert \Phi
_{1}\right\rangle \left\langle \Phi _{1}\right\vert +\left\vert \Phi
_{4}\right\rangle \left\langle \Phi _{4}\right\vert \right]  \nonumber \\
&&+\frac{(A-B)^{2}}{2}\left( \left\vert \Phi _{2}\right\rangle \left\langle
\Phi _{2}\right\vert +\left\vert \Phi _{3}\right\rangle \left\langle \Phi
_{3}\right\vert \right)  \nonumber \\
&&-\frac{(A+B)^{2}}{2}\exp \left( -\left( \mu _{mn}-a_{mn}\right) ^{2}\Gamma
(t)\right)  \nonumber \\
&&\times \left[ \exp \left( -i2(\mu _{mn}-a_{mn})t\right) \left\vert \Phi
_{1}\right\rangle \left\langle \Phi _{4}\right\vert +\exp \left( i2(\mu
_{mn}-a_{mn})t\right) \left\vert \Phi _{4}\right\rangle \left\langle \Phi
_{1}\right\vert \right]  \nonumber \\
&&+\frac{(A-B)^{2}}{2}\exp \left( -\left( \mu _{mn}+a_{mn}\right) ^{2}\Gamma
(t)\right)  \nonumber \\
&&\times \left[ \exp \left( i2(\mu _{mn}+a_{mn})t\right) \left\vert \Phi
_{2}\right\rangle \left\langle \Phi _{3}\right\vert +\exp \left( -i2(\mu
_{mn}+a_{mn})t\right) \left\vert \Phi _{3}\right\rangle \left\langle \Phi
_{2}\right\vert \right]  \nonumber \\
&&+\frac{(A^{2}-B^{2})}{2}\exp \left( -\mu _{mn}^{2}\Gamma (t)\right) 
\nonumber \\
&&\times \left[ \exp \left[ -i\left( 2\mu _{mn}t-\varphi (t)\right) \right]
\left\vert \Phi _{1}\right\rangle \left\langle \Phi _{2}\right\vert +\exp %
\left[ i\left( 2\mu _{mn}t-\varphi (t)\right) \right] \left\vert \Phi
_{2}\right\rangle \left\langle \Phi _{1}\right\vert \right.  \nonumber \\
&&\left. -\exp \left[ -i\left( 2\mu _{mn}t+\varphi (t)\right) \right]
\left\vert \Phi _{3}\right\rangle \left\langle \Phi _{4}\right\vert -\exp %
\left[ i\left( 2\mu _{mn}t+\varphi (t)\right) \right] \left\vert \Phi
_{4}\right\rangle \left\langle \Phi _{3}\right\vert \right]  \nonumber \\
&&+\frac{(A^{2}-B^{2})}{2}\exp \left( -a_{mn}^{2}\Gamma (t)\right)  \nonumber
\\
&&\times \left[ \exp \left[ i\left( 2a_{mn}t+\varphi (t)\right) \right]
\left\vert \Phi _{1}\right\rangle \left\langle \Phi _{3}\right\vert +\exp %
\left[ -i\left( 2a_{mn}t+\varphi (t)\right) \right] \left\vert \Phi
_{3}\right\rangle \left\langle \Phi _{1}\right\vert \right.  \nonumber \\
&&\left. -\exp \left[ i\left( 2a_{mn}t-\varphi (t)\right) \right] \left\vert
\Phi _{2}\right\rangle \left\langle \Phi _{4}\right\vert -\exp \left[
-i\left( 2a_{mn}t-\varphi (t)\right) \right] \left\vert \Phi
_{4}\right\rangle \left\langle \Phi _{2}\right\vert \right] ,  \label{eq9}
\end{eqnarray}%
where $\varphi (t)=4\mu _{mn}a_{mn}C(t)$. The state operator 
$\widehat{\rho }_{IS}(t)$ contains information about the
decay of coherences due to coupling of the system with the environment. 

\section{Decoherence and GHZ state generation probability}

 \begin{figure}[t]
\centering
\includegraphics[width=3.75in,height=5.0in,angle=-90]{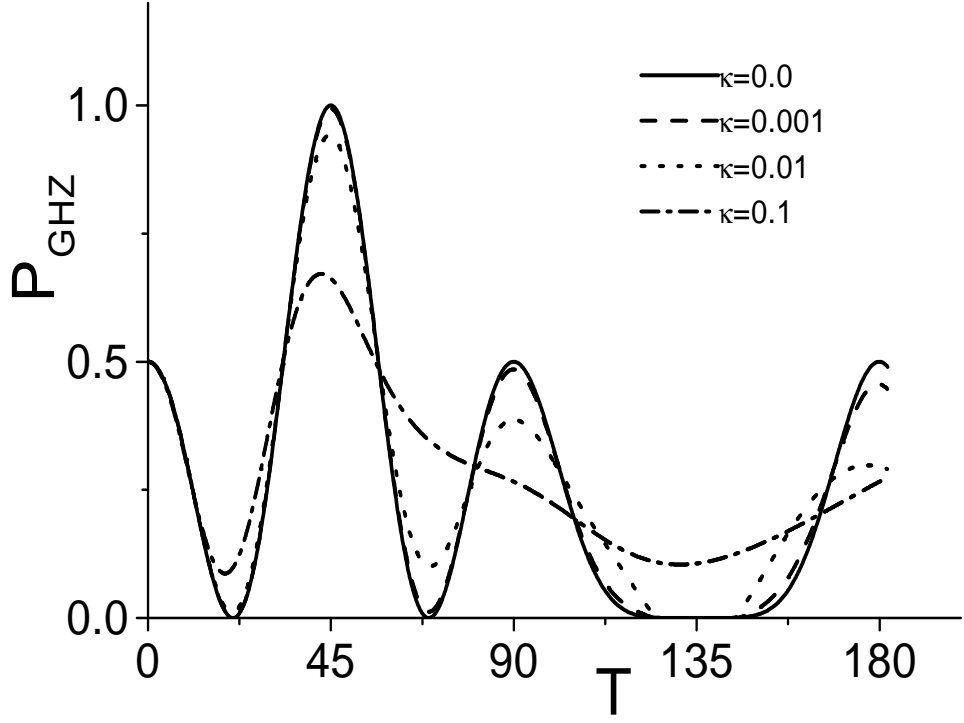}
\caption{ $P_{GHZ}(t)$ as a function of scaled time variable $T(=a_{11}t)$, for
the choice $\mu _{11}\backslash a_{11}=4,$ and $\kappa =0,0.001,0.01$
and $0.1$.}
\label{fig1}
\end{figure}

Undergoing decoherence free time evolution, for  
$\widehat{\rho }_S(0)=\left\vert g,0,0\right\rangle \left\langle
g,0,0\right\vert $ and the interaction time $t_{_{p}}$ such that $%
a_{11}t_{p}=\frac{\pi }{4},\mu _{11}t_{p}=\pi $ ($\alpha =4)$, the system
evolves into a maximally entangled tripartite two mode GHZ state given by
Eq. (\ref{eq232})$.$ We define GHZ state generation probability as $%
P_{GHZ}(t)=tr\left( \widehat{\rho}_{IS} (t)\widehat{\rho }_{GHZ,I}^{(-)}\right) $, 
where $%
\widehat{\rho }_{GHZ,I}^{(-)}=$ $\left\vert \Psi _{GHZ,I}^{-}\right\rangle
\left\langle \Psi _{GHZ,I}^{-}\right\vert $\ . Using Eq. (\ref{eq9})
we write down the density operator $\widehat{\rho }_{IS}(t)$ for the choice $%
m=1,n=1$ and evaluate $P_{GHZ}(t)$. The resulting expression for GHZ\ state
generation probability is 
\begin{eqnarray}
P_{GHZ}(t) &=&\frac{1}{2}+\frac{\Omega ^{2}}{4\mu _{11}^{2}}\left( \exp
\left( \frac{-\mu _{11}^{2}\Gamma (t)}{4.0}\right) \cos \left( 2\mu
_{11}t\right) \cos (\varphi (t))\right)  \nonumber \\
&&+\frac{\Omega ^{2}}{4\mu _{11}^{2}}\left( \exp \left( \frac{%
-a_{11}^{2}\Gamma (t)}{4.0}\right) \sin \left( 2a_{11}t\right) \cos (\varphi
(t))-1\right)  \nonumber \\
&&+\frac{1}{2}\left( \frac{\mu _{11}-a_{11}}{2\mu _{11}}\right) ^{2}\exp
\left( \frac{-\left( \mu _{11}-a_{11}\right) ^{2}\Gamma (t)}{4.0}\right)
\sin \left[ 2(\mu _{11}-a_{11})t\right]  \nonumber \\
&&-\frac{1}{2}\left( \frac{\mu _{11}+a_{11}}{2\mu _{11}}\right) ^{2}\exp
\left( \frac{-\left( \mu _{11}+a_{11}\right) ^{2}\Gamma (t)}{4.0}\right)
\sin \left[ 2(\mu _{11}+a_{11})t\right] .  \label{eq11}
\end{eqnarray}%
We also calculate the population inversion defined as $I=P_{g}-P_{e},$ where 
$P_{g}(P_{e})$ is the probability of finding the ion in ground (excited)
state. For the choice $m=1,n=1,$ in the state $\widehat{\rho }_{IS}(t)$ of Eq. (%
\ref{eq9}), we get 
\begin{eqnarray}
I &=&\frac{\mu _{11}+a_{11}}{2\mu _{11}}\left( \exp \left( -\left( \mu
_{11}-a_{11}\right) ^{2}\Gamma (t)\right) \cos \left[ 2(\mu _{11}-a_{11})t%
\right] \right)  \nonumber \\
&&+\frac{\mu _{11}+a_{11}}{2\mu _{11}}(\exp \left( -\left( \mu
_{11}+a_{11}\right) ^{2}\Gamma (t)\right) \cos \left[ 2(\mu _{11}+a_{11})t%
\right] . \label{eq12}
\end{eqnarray}

In the limit $\kappa \rightarrow 0$ the Eqs. (\ref{eq9}), (\ref{eq11}) and (%
\ref{eq12}) reduce to decoherence free evolution of the state operator, the
GHZ\ state generation probability, and the population inversion,
respectively. As the cavity is initially in vacuum state and ionic center of
mass in the lowest energy trap state, decoherence effects due to cavity
decay and heating are expected to be small and have not been considered.

\begin{figure}[t]
\centering
\includegraphics[width=3.75in,height=5.0in,angle=-90]{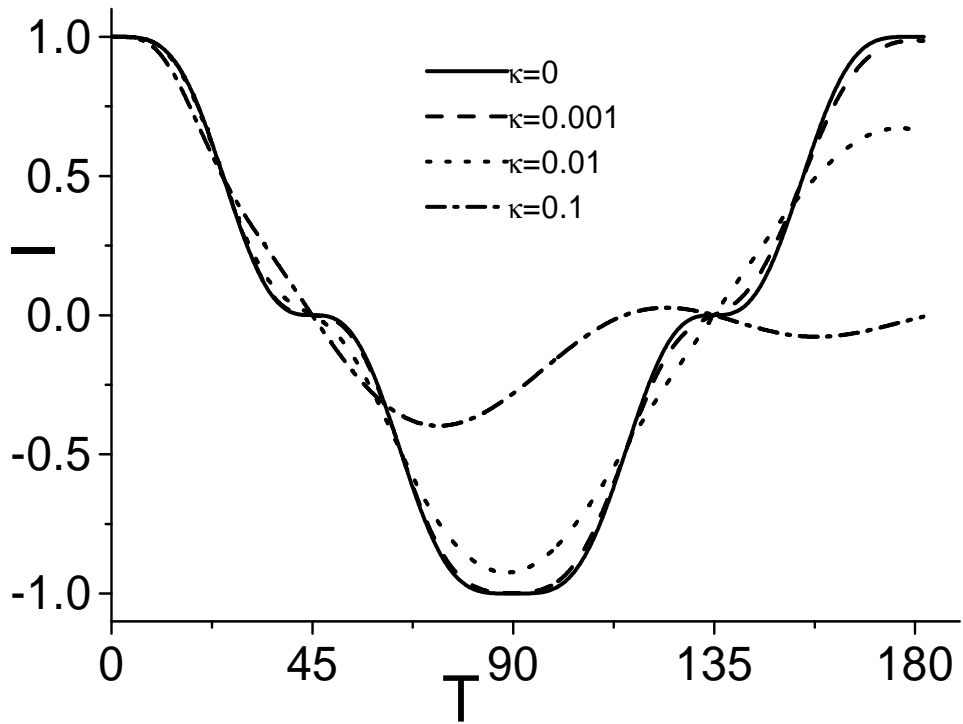}
\caption{ Population inversion $I(t),$ as a function of scaled time variable $%
T(=a_{11}t)$, for the choice $\mu _{11}\backslash a_{11}=4,$ and $\kappa
=0, 0.001, 0.01$ and $0.1$.}
\label{fig2}
\end{figure}

\section{Negativity and Linear entropy}

The state of three qubit composite system is represented by $\widehat{%
\rho }_{IS}(t)=\widehat{\rho }_{ABC}(t),$ where $A$, $B$, and $C$ refer to
the internal state of two-level ion, the state of vibrational motion, and
the cavity state respectively. A pure state of a tripartite system can be
separable, or entangled. Entangled state may or may not have(biseparable)
true tripartite entanglement. Bipartite entanglement is relatively well
understood. Entanglement content of a tripartite system may be measured
through the entanglement of its bipartite decompositions that is A+(BC),
B+(AC), and C+(AB). The decoherence effects result in a composite state $%
\widehat{\rho }_{ABC}(t),$ which is usually a mixed state. Mixed state
entanglement is less understood than the pure state entanglement. On
examining Eq. (\ref{eq9}), in which the state operator is expressed in the
eigen-basis of $\hat{H}_{IS}$, we notice that the diagonal matrix elements
of density matrix are not affected by coupling to the environment, whereas
the moduli of off diagonal matrix elements decay with time. Consequently,
when only non-dissipative decoherence is present the trace of $\widehat{\rho 
}_{ABC}(t)$ remains constant with time and the eigenvalues continue to be
positive. Since $\widehat{\rho }_{ABC}(t)$ is positive definite, we use
negativity proposed by Vidal and Werner \cite{vida02} as an entanglement
measure for pure as well as mixed states. Negativity is based on partial
transpose of density matrix of composite system with respect to a subsystem.

The state of a bipartite system, composed of subsystems $A$ and $B$ in 
finite dimensional Hilbert spaces $d_{A}$ and $d_{B}$, is written as
\begin{equation}
\widehat{\rho }=\sum_{i,j=1}^{d_{A}}\sum_{m,r=1}^{d_{B}}\left\langle
i,m\left| \widehat{\rho }\right| j,r \right\rangle \left| \
i,m\right\rangle \left\langle j,r \right| .  \label{3.0}
\end{equation}
The partial transpose of density operator with respect to
sub-system A is defined as
\begin{equation}
\widehat{\rho }^{T_{A}}=\sum_{i,j=1}^{d_{A}}\sum_{m,r}^{d_{B}}\left\langle
i,m\left| \widehat{\rho }\right| \ j,r\right\rangle \left| \
j,m\right\rangle \left\langle \ i,r\right| .  \label{3.1}
\end{equation}
The partial transpose of density matrix of an entangled state is not
positive definite. Negativity defined as {\sl N}$^{A}=\sum_{i}\left| \lambda
_{i}\right| ,$ where $\lambda _{i}$ are the negative eigenvalues of \ $%
\widehat{\rho }^{T_{A}}$ is a measure of entanglement of quantum system $A$
with $B$. For a separable state ${N}^{A}=0$ and at maximal entanglement
value of ${N}^{A}$ depends on dimension $d_{A}.$ For the tripartite system
at hand, we construct the partial transpose with respect to sub-systems $A$, 
$B$ or $C$, while keeping the composite state of the remaining two
sub-systems unaltered. For the density matrix operator $\widehat{\rho }%
_{ABC}(t)$ acting on composite space, the transpose with respect to
sub-system A reads as 
\begin{equation}
\widehat{\rho }^{T_{A}}=\sum_{i,j=1}^{d_{A}}\sum_{m,r=1}^{d_{B}}%
\sum_{n,s=1}^{d_{C}}\left\langle i,m,n\left| \widehat{\rho }\right| \
j,r,s\right\rangle \left| \ j,m,n\right\rangle \left\langle \ i,r,s\right| ,
\label{3.3}
\end{equation}
where $d_{X}(X=A,B,C)$ refers to dimension of the Hilbert space of subsystem 
$X$. We use $\widehat{\rho }^{T_{A}}$, $\widehat{\rho }^{T_{B}}$, and 
$\widehat{\rho }^{T_{C}}$ to calculate the negativitiy measures for
sub-systems $A$, $B$, and $C$, respectively 

\begin{figure}[t]
\centering
\includegraphics[width=3.75in,height=5.0in,angle=-90]{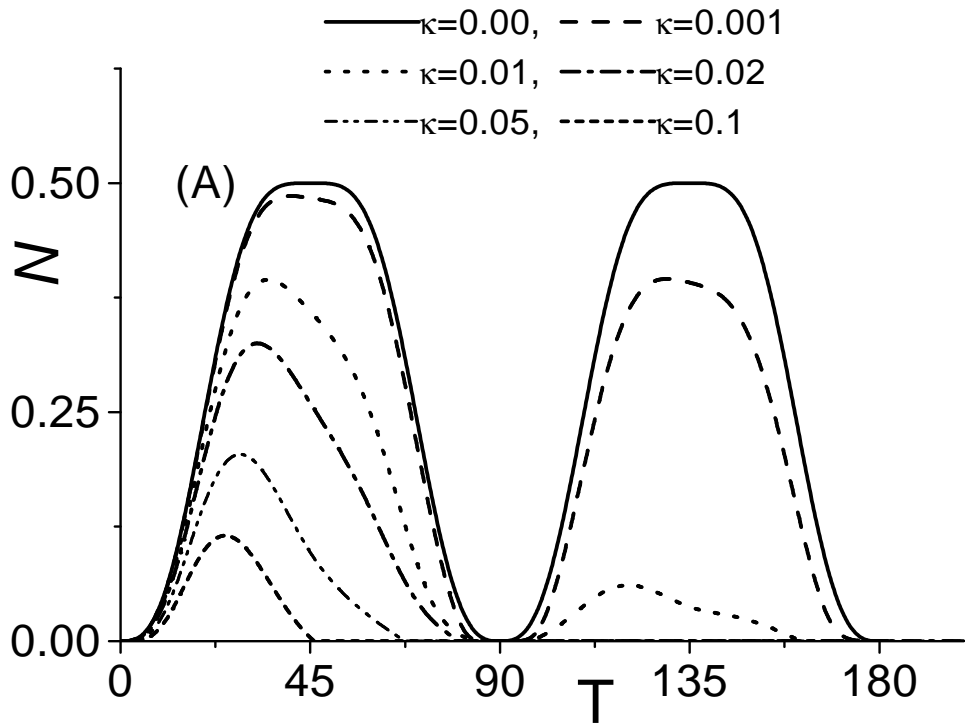}
\caption{ \textsl{N}, calculated from $\widehat{\rho }%
^{T_{A}}$, 
as a function of scaled time variable $T(=a_{11}t)$, for the choice $\mu
_{11}\backslash a_{11}=4,$ and $\kappa =0, 0.001, 0.01, 0.02, 0.05$ 
and $0.1$.}
\label{fig3}
\end{figure}

\begin{figure}[t]
\centering
\includegraphics[width=3.75in,height=5.0in,angle=-90]{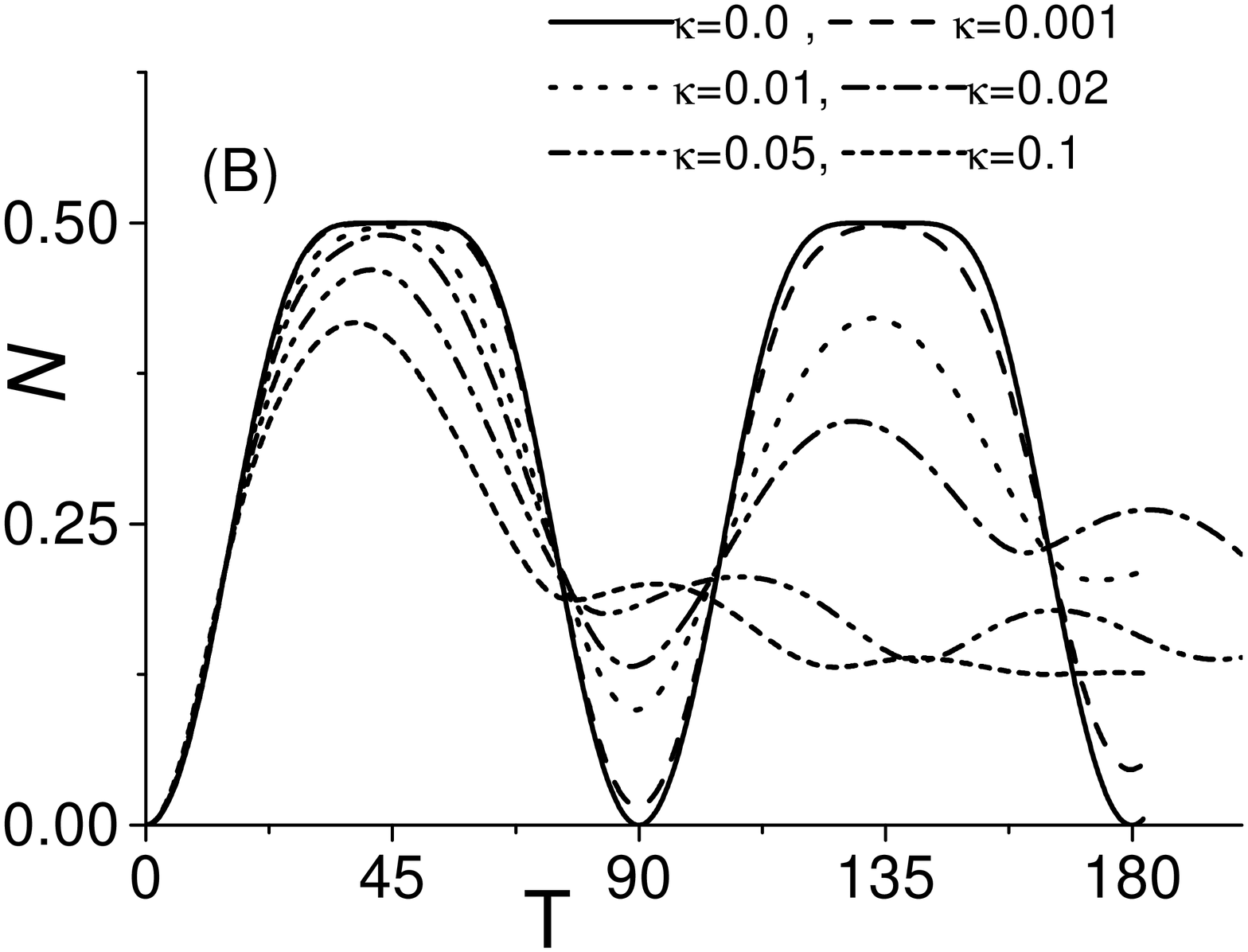}
\caption{ \textsl{N}, calculated from 
$\widehat{\rho }^{T_{B}}=\widehat{\rho }^{T_{C}}$, as a
function of scaled time variable $T(=a_{11}t)$, for the choice $\mu
_{11}\backslash a_{11}=4,$ and $\kappa =0, 0.001, 0.01, 0.02, 0.05$
 and $0.1$.}
\label{fig4}
\end{figure}

For a quantum state $\widehat{\rho }$ in $d$ dimensional Hilbert space, the
linear entropy $S_{l}$ is defined as 
\begin{equation}
S_{l}=\frac{d}{d-1}\left(1-Tr ( \widehat{\rho }^{2})\right ).  \label{3.4}
\end{equation}
Linear entropy is used to measure the mixedness of subsystems 
$A,B,$ and $C$. For pure states $S_{l}=0,$
whereas for maximally mixed states $S_{l}=1.$ Reduced state operators $%
\widehat{\rho }_{red}^{A}=tr_{BC}(\widehat{\rho }_{ABC})$, $\widehat{\rho }%
_{red}^{B}=tr_{AC}(\widehat{\rho }_{ABC})$, and $\widehat{\rho }_{red}^{C}$ $%
=tr_{AB}(\widehat{\rho }_{ABC})$ are used to calculate linear entropy $S_{l}$,
for the subsystems $A$, $B$, and $C$, respectively. Negativities and linear 
entropies are used to characterize the state of the composite system at 
an instant t, for a given value of coupling to the environment.
\begin{figure}[t]
\centering
\includegraphics[width=3.75in,height=5.0in,angle=-90]{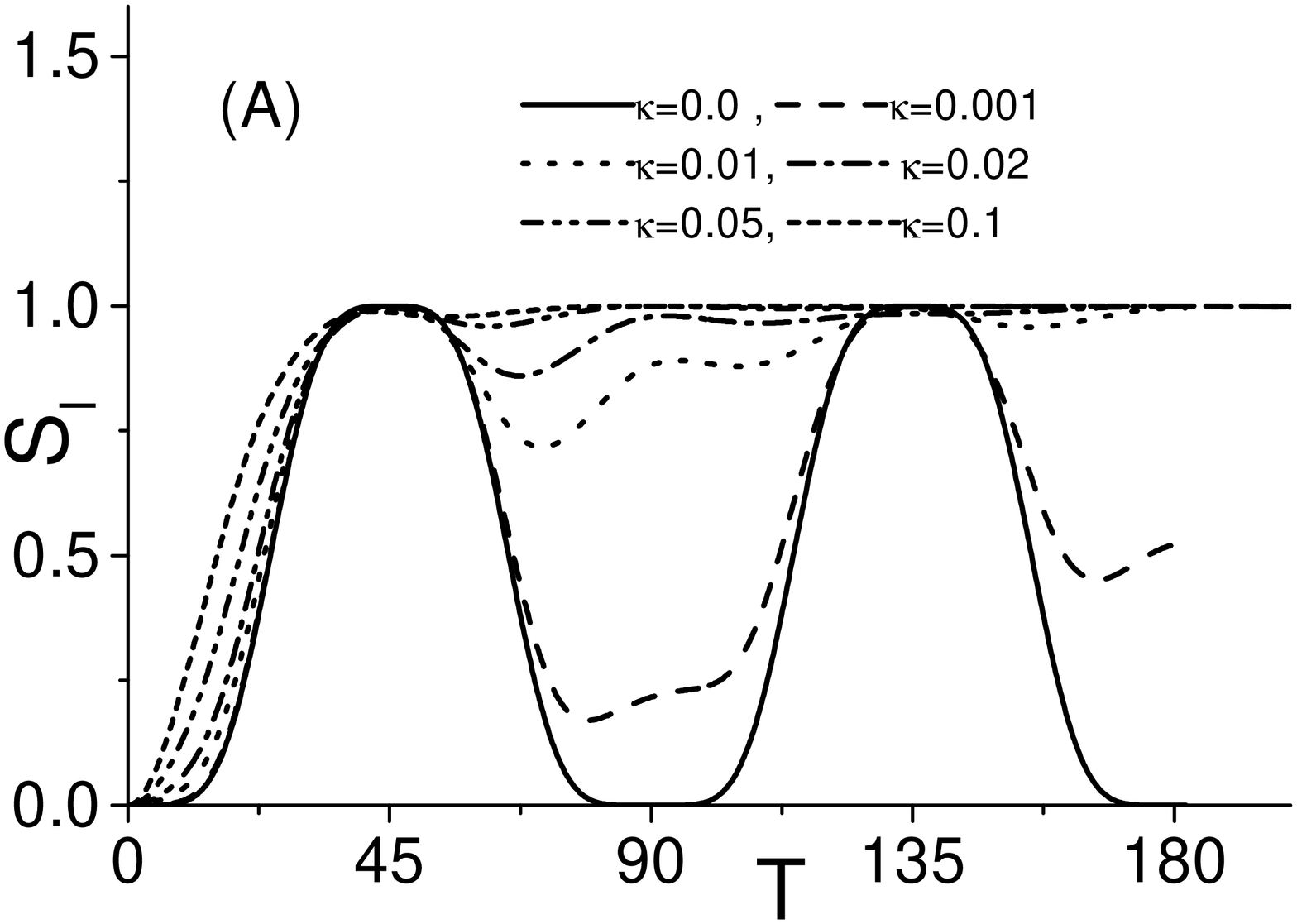}
\caption{ Linear entropy \textsl{S}$_{l}^{A}$, calculated from 
$\widehat{\rho }_{red}^{A}$, as a function of scaled time 
variable $T(=a_{11}t)$, for
the choice $\mu _{11}\backslash a_{11}=4,$ and 
$\kappa =0, 0.001, 0.01, 0.02, 0.05$
 and $0.1$.}
\label{fig5}
\end{figure}

\section{Numerical results and conclusions}

We consider the system prepared in the state $\widehat{\rho }_{S}(0)=\left|
g,0,0\right\rangle \left\langle g,0,0\right| $, at $t=0.$ The laser-ion
coupling constant is taken to be $\Omega =8.95$ $MHz$ \cite{mund03}$,$ where
as the ratio $\mu _{11}/a_{11}=4.0$. Numerical values of $\Gamma (t)$ and $%
C(t)$ are obtained by solving the integrals of Eqs. (\ref{eq8a},\ref{eq8b}) for
the choice $\omega _{c}=1200MHz,$ temperature $T=0.03K$ and different values of
coupling strength $\kappa .$ The variable $\omega $ in 
Eqs. (\ref{eq8a},\ref{eq8b}) is varied from zero to $%
3\omega _{c}.$ The value $\kappa =0.001$ corresponds to a weak coupling and
successively larger values tend to produce decoherence on a scale comparable
to GHZ state generation time which is of the order of $0.34$ nano s$.$ The
state of the system at an instant $t$ can be detected experimentally by
cavity-photon measurement combined with atomic population inversion
measurement.

Figs. (1) and (2) display $P_{GHZ}(t)$ and population inversion $I(t),$
respectively, as a function of scaled time variable $T(=a_{11}t)$ for
decoherence parameter values of $\kappa =0,0.001,0.01,$ and $0.1$. Evidently
the evolution dynamics of $P_{GHZ}(t)$ and $I(t)$ is sensitive to changes in
the decoherence parameter $\kappa $. The presence of decay factors in Eq. (%
\ref{eq11}) causes the peak value of $P_{GHZ}$ to decrease with time. Phase
decoherence causes a shift in the interaction time needed to get the system
in maximally entangled state. The population inversion however does not show
any phase decoherence effects. We may point out that at $T=135^{\circ }$ %
the tripartite system is in maximally entangled state of Eq. (\ref%
{eq233}) which is orthogonal to the state of Eq. (\ref{eq232})for 
which $P_{GHZ}(t)$ has been
calculated. Population inversion is zero at $T=45^{\circ }$ as well as $%
135^{\circ },$ signalling maximally entangled state generation. For weak
coupling ($\kappa =0.001$), the behavior of $I(t)$ does not differ much from
no decoherence curve. However, for larger values of $\kappa $, $I(t)$ for the
separable system at $T=45^{\circ }$ can indicate the amount of decoherence
present.

The state operator of Eq. (\ref{eq9}) for $m=1,n=1$ is used to calculate the 
density matrices transposed with respect to quantum systems A, B, and C. 
The negativities are calculated, numerically, from $\widehat{\rho }^{T_{A}}$, 
$\widehat{\rho }^{T_{B}}$, and $\widehat{\rho }^{T_{C}}$%
and plotted as a function of scaled time variable $T(=a_{11}t)$. 
The negativity plot displaying entanglement dynamics of
internal state of two-level ion ($A$) is shown in Fig. 3. For subsystem $B $, the 
negativity calculated from $\widehat{\rho }^{T_{B}}$=$\widehat{\rho }^{T_{C}}$, is 
displayed as a function $T(=a_{11}t)$ in Fig. 4.
The values of
decoherence parameter are $\kappa =0,0.001,0.01,0.02,0.05$ and $0.1$. The
time evolution of vibrational motion of ion's center of mass($B$) and cavity
number state ($C$) are identical due to inherent symmetry of these subsystems. 
We notice that as the value of decoherence parameter increases, there is a
decrease in the maximum entanglement of sub-systems $A$, $B$ and $C$ with their
compliments. Besides that the phase decoherence changes the time at which
maximal entanglement of the three parties is observed. These two factors are
consistent with the behavior of\ $P_{GHZ}(t)$ and $I(t)$ in Figs. 1 and 2.

Linear entropies $S_{l}$, for subsystems $A$, $B$ and $C$ have been 
calculated using the reduced density operators obtained from $\widehat{\rho }%
_{IS}(t)$ of Eq. (\ref{eq9}) for $m=1,n=1$. Figs. 5 and 6, display $S_{l}$ 
as a function of scaled time variable $T(=a_{11}t)$, for subsystems 
$A$ and $B$(or $C$), respectively. 
The decoherence parameter values are $\kappa
=0,0.001,0.01,0.02,0.05$ and $0.1$. For $\kappa =0.001,0.01,$ the trend of
time evolution of linear entropy follows the variation of entanglement as
seen in negativity plots of Figs. 3 and 4. However, for a stronger coupling
value of $\kappa =0.02$ for subsystem $A$, whereas {\sl N} tends to zero for $T\ $%
approaching $90^{0},$ $S_{l}$  shows fluctuation for $T>90^{0}$. 
Similarly for $\kappa =0.05$ negativity indicates a decoupling of system $A$
from system $BC$ beyond $T=67^{0}$ while $S_{l}$ points to the
presence of some correlations. For $\kappa =0.1,$ the system is highly damped
and both {\sl N} and $S_{l}$ show the ion to become separable
around $T=45^{0}$. We may conclude that for the states lying at the
boundary of entangled separable region the negativity and linear entropy show 
distinctly different variation with time the reason being that the negativity 
measures only entanglement generating quantum correlations, whereas the linear 
entropy measures all correlations that reduce the purity of the state.

\begin{figure}[t]
\centering
\includegraphics[width=3.75in,height=5.0in,angle=-90]{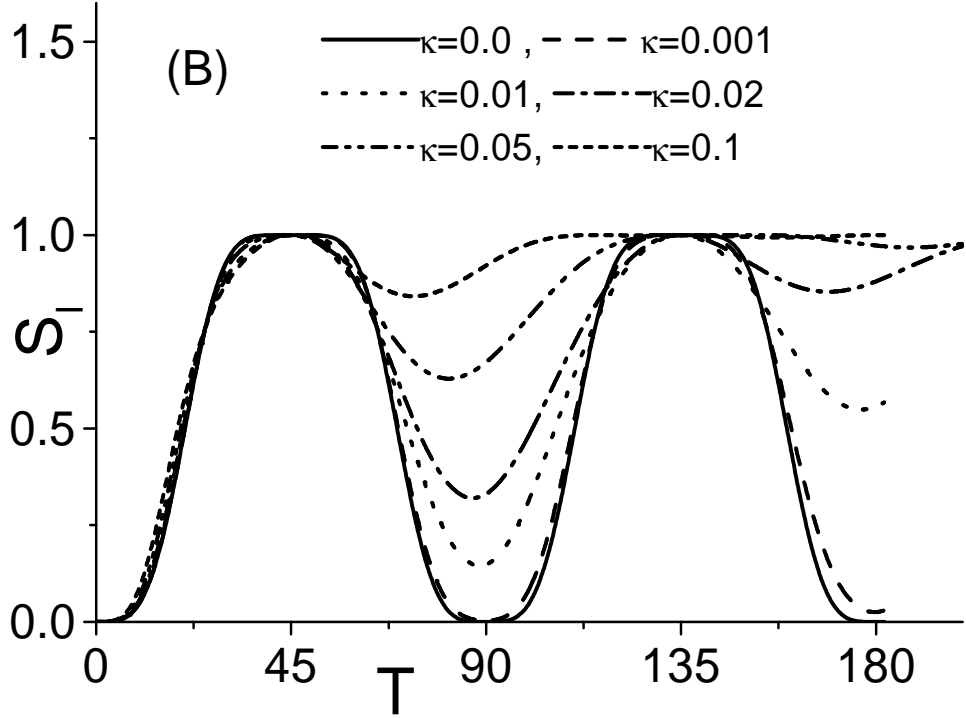}
\caption{ Linear entropy \textsl{S}$_{l}$, calculated from 
$\widehat{\rho }_{red}^{B}$,
as a function of scaled time variable $T(=a_{11}t)$, for the choice $\mu
_{11}\backslash a_{11}=4,$ and $\kappa =0, 0.001, 0.01, 0.02, 0.05$
 and $0.1$.}
\label{fig6}
\end{figure}

To conclude, we have investigated the decoherence process of three qubit
system obtained by manipulating the state of a cold trapped two-level ion
coupled to an optical cavity. Interaction of the ion with a resonant laser
field and the cavity field tuned to red sideband of ionic vibrational
motion, generates entanglement of internal states of the ion, vibrational
states of ionic center of mass and the cavity field state. Non-dissipative
decoherence of the state of the system occurs due to interaction of the
system with the environment. We have considered the effect of decoherence 
sources such as magnetic field fluctuations, laser frequency fluctuations, 
potentials applied 
to elctrodes and coupling to states outside the model space. The random 
perturbations of the Hamiltonian are accounted for by coupling the model space
interaction Hamiltonian to the the environment, modeled as a set of
non-interacting harmonic oscillators. The pointer observable is energy of
the isolated quantum system. Analytic expression for the state operator of
tripartite composite system, including non dissipative decoherence effects 
has been obtained. The initial state of
the tripartite system is a separable state and the state of environment has
no initial correlations. Coupling to environment is found to introduce an
exponential decay with time of the off diagonal matrix elements of density
operator $\widehat{\rho} _{IS}(t)$. In addition, a defasing of various states
in the superposition also occurs. The extent to 
which the process is decohered depends on the system envionment
coupling strenth, that can be partially controlled by adjusting the 
electrode potentials. Besides the strength of coupling to the
environment, the energy spectrum of system Hamiltonian plays an important
role in determining the decoherence rate of a given initial state of the
composite system. For a specific choice of interaction parameters, the
isolated system evolves to tripartite GHZ state \cite{shar103}. We have
evaluated analytically the probability of generating maximally entangled GHZ
state, and the population inversion in the presence of non-dissipative
decoherence. Numerical calculations for different values of system
environment coupling strengths using interaction parameter from a recent
experiment show the peak value of GHZ state generation probability\ to
decrease with increase in $\kappa .$ $P_{GHZ}(t)$ is also sensitive to phase
decoherence. The population inversion however does not show any phase
decoherence effects. The results can be used to study the effects of 
engineered environment on the process of tripartite maximally entangled 
state generation. Bipartite entanglement of cavity mode and atomic internal 
states can be produced without coupling to the motional states. The tripartite 
entanglement generated through coupling to motional states can be transferred 
to other ions that might be added to the trap. In comparison with an earlier 
proposal for GHZ state generation \cite{fidi02}, in the current scheme the initial 
state of the system is $\left\vert g,0,0\right\rangle$. As such the cavity 
losses and spontaneous decay effects are minimized.  

Using negativity as an entanglement measure, the entanglement dynamics of
the tripartite system in the presence of decoherence has been analyzed. As
the value of decoherence parameter increases, the maximum entanglement of
sub-systems $A$ and $B$ with their compliments decreases. In the context of
trapped ion radiated by the single mode cavity field and an external
resonant laser, the time for which the two level ion remains coupled to the
state of vibrational motion and the cavity state decreases with increase in $%
\kappa .$ For large values of $\kappa $ the states of the composite system
lying at the boundary of entangled-separable region are reached. For these 
states the negativity and linear entropy show distinctly 
different variation with time. This can be understood by noting that whereas 
the negativity measures entanglement generating quantum correlations, the linear 
entropy measures all correlations that reduce the purity of the state.

{\Large{Acknowledgments}}

S. S. Sharma and N. K. Sharma acknowledge financial support from CNPq, Brazil. 
E. de Almeida thanks Capes, Brazil for financial support.

\end{document}